\documentclass[twocolumn,twocolappendix]{aastex631}

\usepackage[T1]{fontenc}
\usepackage{ae,aecompl}
\usepackage{multirow}
\usepackage{url}
\usepackage{enumitem}

\usepackage{graphicx}	
\usepackage{amsmath}	
\usepackage{amssymb}	
\usepackage{float}
\usepackage[caption=false]{subfig}

\defcitealias{mauerhofer2021}{M21}

\newcommand{\cII}{\ion{C}{2}}

\newcommand{\lya}{Ly$\alpha$}

\newcommand{\cIIl}{\ion{C}{2}~$\lambda1334$}
\newcommand{\cIIlstar}{\ion{C}{2}*~$\lambda1335$}

\newcommand{\siII}{\ion{Si}{2}}
\newcommand{\siIIl}{\ion{Si}{2}~$\lambda1260$}
\newcommand{\siIIlstar}{\ion{Si}{2}*~$\lambda1265$}

\newcommand{\classy}{\textsc{classy}}
\newcommand{\vandels}{\textsc{vandels}}
\newcommand{\fesclyc}{$f_{\rm esc}^{\rm LyC}$}
\newcommand{\fesclycpre}{$f_{\rm esc\ (pred)}^{\rm LyC}$}

\newcommand{\fesclya}{$f_{\rm esc}^{\rm Ly\alpha}$}

\begin{document}


\title{Comparing the VANDELS sample to a zoom-in Radiative Hydrodynamical Simulation: using the Si II and C II line spectra as tracers of galaxy evolution and Lyman Continuum leakage}

\author[0000-0002-5659-4974]{Simon Gazagnes}
\affiliation{Department of Astronomy, The University of Texas at Austin, 2515 Speedway, Stop C1400, Austin, TX 78712-1205, USA}

\author[0000-0002-3736-476X]{Fergus Cullen}
\affiliation{Institute for Astronomy, University of Edinburgh, Royal Observatory, Edinburgh EH9 3HJ}

\author[0000-0003-0595-9483]{Valentin Mauerhofer}
\affiliation{Kapteyn Astronomical Institute, University of Groningen, P.O Box 800, 9700 AV Groningen, The Netherlands}

\author[0000-0003-0629-8074]{Ryan Begley}
\affiliation{Institute for Astronomy, University of Edinburgh, Royal Observatory, Edinburgh EH9 3HJ}

\author[0000-0002-4153-053X]{Danielle Berg}
\affiliation{Department of Astronomy, The University of Texas at Austin, 2515 Speedway, Stop C1400, Austin, TX 78712-1205, USA}

\author[0000-0003-1609-7911]{Jeremy Blaizot}
\affiliation{Centre de Recherche Astrophysique de Lyon, UMR5574, F-69230, Saint-Genis-Laval, France}

\author[0000-0002-0302-2577]{John Chisholm}
\affiliation{Department of Astronomy, The University of Texas at Austin, 2515 Speedway, Stop C1400, Austin, TX 78712-1205, USA}

\author[0000-0002-9613-9044]{Thibault Garel}
\affiliation{Department of Astronomy, University of Geneva, 51 Chemin Pegasi, 1290 Versoix, Switzerland}

\author[0000-0002-6085-5073]{Floriane Leclercq}
\affiliation{Department of Astronomy, The University of Texas at Austin, 2515 Speedway, Stop C1400, Austin, TX 78712-1205, USA}

\author{Ross J. McLure}
\affiliation{Institute for Astronomy, University of Edinburgh, Royal Observatory, Edinburgh EH9 3HJ}

\author[0000-0002-2201-1865]{Anne Verhamme}
\affiliation{Department of Astronomy, University of Geneva, 51 Chemin Pegasi, 1290 Versoix, Switzerland}



\begin{abstract}

We compare mock ultraviolet \cII\ and \siII\ absorption and emission line features generated using a $\sim10^9$ $M_\odot$ virtual galaxy with observations of 131 $z\sim3$ galaxies from the \vandels\ survey.  We find that the mock spectra reproduce reasonably well a large majority (83\%) of the \vandels\ spectra ($\chi^2<2$), but do not resemble the most massive objects ($\gtrapprox10^{10}$M$_\odot$) which exhibit broad absorption features. Interestingly, the best-matching mock spectra originate from periods of intense star formation in the virtual galaxy, where its luminosity is four times higher than in periods of relative quiescence. Furthermore, for each galaxy, we predict the Lyman Continuum (LyC) escape fractions (\fesclycpre) using the environment of the virtual galaxy. We derive an average \fesclycpre\ of 0.01$\pm$0.02, consistent with other estimates from the literature. The \fesclycpre\ are tightly correlated with the Lyman-$\alpha$ escape fractions and highly consistent with observed empirical trends. Additionally, galaxies with larger \fesclycpre\ exhibit bluer $\beta$ slopes, more Lyman-$\alpha$ flux, and weaker low-ionization absorption lines.  Building upon the good agreement between \fesclycpre\ and observationally established LyC diagnostics, we examine the LyC leakage mechanisms in the simulation. We find that LyC photon leakage is enhanced in directions where the observed flux dominantly emerges from compact regions depleted of neutral gas and dust, mirroring the scenario inferred from observational data. In general, this study further highlights the potential of high-resolution radiation hydrodynamics simulations in analyzing UV absorption and emission line features and providing valuable insights into the LyC leakage of star-forming galaxies.
\end{abstract}







\section{Introduction} \label{sec:intro}

Ultraviolet (UV) absorption and emission line features are formidable tracers of the Interstellar and Circumgalactic Media (ISM, CGM) properties within star-forming galaxies. These UV features provide direct insights into key gas properties, including its metallicity, geometry, and kinematics. Such parameters are fundamental in tracking the evolving environment of galaxies over cosmic epochs, forming a cornerstone for the establishment of precise galaxy evolution models. In particular, the analysis of UV features enables us to explore properties intrinsically linked to the baryon cycle of galaxies, quantify the metal content of neutral gas, and infer the metal enrichment of the ISM \citep[e.g.,][]{heckman2000, James2014, heckman2015, shapley2003, erb2012, rubin2011, Wofford2013, chisholm2015, chisholm2017mass,  finley2017wind, steidel2018, wang2020, Xu2022}. 

UV absorption lines originating from low-ionization states (LIS) of metals, such as \cII\ or \siII, hold particular relevance in the context of reionization investigations. These absorption features, characterized by substantial residual flux (i.e., the flux remaining at the line's maximum depth) or relatively low equivalent widths (EWs), have been identified in various galaxy samples exhibiting ionizing photon leakage \citep[referred to as Lyman Continuum Emitters or LCEs, ][]{reddy2016stack, gazagnes2018, saldana2022}. Indeed, these features directly probe the absence of dense neutral clouds or the presence of low-column density channels through which ionizing photons can escape the galaxy's ISM \citep[e.g.,][]{heckman2001, alexandroff2015, reddy2016stack, gazagnes2018, chisholm2018, steidel2018, saldana2022}. This property underscores the potential of LIS absorption lines as promising diagnostic tools for predicting the escape fraction of ionizing photons ($f_{\rm esc}^{\rm LyC}$, where LyC represents the Lyman Continuum photons with $\lambda<912$\AA) in the high-redshift Universe, where direct measurements are hindered by the opacity of the intergalactic medium (IGM).

While UV LIS absorption and emission lines offer substantial potential for exploring the characteristics of a galaxy's ISM and constraining the LyC photons escaping star-forming galaxies, conventional analyses and diagnostics of these features often rely on spectrally averaged measurements, such as equivalent width or residual flux, which merely capture a fraction of the rich informational potential. In practice, the emergent profiles of UV absorption and emission lines can exhibit a high degree of complexity resulting from the interplay of multiple gas clouds. The ISM geometry and complexity are critical keys to understanding the physical processes that have transformed the galaxy environment and might have enabled the leakage of LyC photons \citep[e.g., ][]{jaskot2019, rivera2017}. To capture the vital features that are directly related to certain properties such as the LyC escape, one must imperatively advance our analytical methodologies to fully exploit the amount of information available within these features.

Over the past decades, semi-analytical techniques building upon outflow and inflow geometrical models have shown promising results for interpreting simultaneously and consistently the resonant and fluorescent features of low and high ionization states of metals \citep[e.g.,][]{scarlata2015, carr2018, carr2022, carr2023_rascas_comp,li2024_alpaca}. Simultaneously, the development and progress of hydrodynamical simulation techniques have made a significant step in the development of new frameworks for the interpretation of spectroscopic observations. Building upon \citet{mauerhofer2021}, whose authors used a zoom-in high-resolution radiation hydrodynamics (RHD) simulation post-processed with radiative-transfer techniques, \citet{blaizot2023simulating} highlighted that the mock Lyman$-\alpha$ (\lya) profiles produced in such a simulation closely resemble some of the notoriously complex \lya\ profiles observed in the low$-z$ universe. \citet{gazagnes2023} showed that the same simulation produced mock \cII\ and \siII\ spectra (including the fluorescent \cII* and \siII* features) which could well reproduce 38 out of 45 observations of low$-z$ galaxies from the COS Legacy Archive Survey \citep[\classy,][]{berg2022}. This strong correspondence between UV absorption and emission lines from star-forming galaxies and simulated spectra generated from a single virtual galaxy has opened a promising avenue for interpreting galaxy observations. It suggests that we can now construct physically meaningful models that faithfully replicate realistic environments and help us connect observations to the actual physical galaxy properties responsible for the emergent spectral properties  \citep[see also][]{Choustikov2023_, Katz2023_sphinxrelease}.


In this work, we extend our exploration of how mock observations from high-resolution simulations can offer a novel and insightful approach to connect spectroscopic observations to the underlying physical processes governing galaxy evolution. Our investigation dives into two distinct aspects. First, while prior work by \citet{blaizot2023simulating} and \citet{gazagnes2023} focused on observations at redshifts around $z < 0.5$, we assess the simulation's reliability in replicating the \cII\ and \siII\ spectra of a more extensive galaxy sample at 3.35 $<$ $z$ $<$ 3.95, drawn from the ultra-deep VANDELS spectroscopic survey \citep{McLure2018_vandels, Pentericci2018_vandels, Garilli2021_vandels}. Secondly, we investigate the origin of the similarities between observed and simulated spectra and employ the simulation as a predictive tool for Lyman Continuum escape (\fesclyc), comparing these predictions with both direct and indirect \fesclyc\ constraints from \citet{Begley2022_lyc} and \citet{saldana2023}, as well as with the \lya\ escape fraction measurements from \citet{Begley2023_lya_lyc}. One objective of this paper is to further illustrate the potential of high-resolution RHD simulations in the interpretation of UV spectroscopic observations and in the extraction of meaningful LyC diagnostics, an aspect that holds particular relevance in the current landscape of high-redshift observations.


This paper is organized as follows: Section~\ref{sec:data} describes the simulation and observation data sets used in this work. Section~\ref{sec:measurement} compares the \cIIl+\cIIlstar\ and \siIIl+\siIIlstar\ spectra in the simulation and in the observations. Section~\ref{sec:galprop} discusses the origin of the resemblance between the mock and observed spectra in the context of the virtual galaxy properties. Section~\ref{sec:lyc} presents  \fesclyc\ predictions made using the simulation and explores their consistency with previous LyC constraints and the shape of UV spectral features. Finally, Section~\ref{sec:sum} summarizes our findings.

\section{Data} \label{sec:data}
In this work, we compare the \cIIl+\cIIlstar\ and \siIIl+\siIIlstar\ observations from \vandels\ to mock spectra generated from the simulated galaxy of \citet{mauerhofer2021}. Section~\ref{sec:datasim} describes the simulation and Section~\ref{sec:dataobs} details the VANDELS observations. 

\subsection{Zoom-in simulation}
\label{sec:datasim}

The RHD simulation used in the work is identical to the one used in \citet{gazagnes2023} where we analyzed the  \cIIl+\cIIlstar\ and \siIIl+\siIIlstar\ observations from the \classy\ sample \citep{berg2022}. Therefore, we only present here the most important characteristics of the simulation and refer the readers to Section 2 of \citet{gazagnes2023} for a complete and thorough description. 

The zoom-in simulation run was originally presented in \citet{mauerhofer2021}, and was obtained with the adaptive mesh refinement code \textsc{Ramses-RT} \citep{teyssier2002, rosdahl2013, rosdahl2015} which builds upon the physics of cooling, supernova feedback, and star-formation from \textsc{sphinx} \citep{rosdahl2018}. The initial conditions were generated using \textsc{music} \citep{Hahn2013} and chosen such that the resulting galaxy has a stellar mass of M$_\star\sim10^9$ M$_\odot$ at $z = 3$, with a maximum cell resolution of 14 pc around this redshift. We selected 75 outputs with 10 Myrs time-steps between $z = 4.19$ to $z = 3.00$. During this period, the halo mass increases from $10^{10.6}$ to $10^{10.75}$~$M_\odot$, the stellar mass ($M_\star$) increases from $10^{8.8}$ to $10^{9.4}$~${\rm M_\odot}$, both the averaged stellar and gas metallicity increase from $0.20Z_\odot$ to $0.42Z_\odot$, and the star-formation rate (SFR) oscillates between $\sim1.5$ and 4.0~$M_\odot$ yr$^{-1}$, with two main SFR peaks around 1.64 and 2.04 Gyrs (the evolution of the SFR is shown and discussed in Section~\ref{sec:spec_ori}). 

To produce the LIS mock spectra, we first compute the number densities of \cII\ and \siII\ using the following procedure: we derive the densities of carbon and silicon in each simulation cell assuming solar abundance ratios, and then compute their ionization fractions with the customizable chemical network code \textsc{krome} \citep{grassi2014}, using the recombination rates from \citep{badnell2006}, the collisional ionization rates from \citep{voronov1997}, the photoionization cross-sections from \citep{verner1996}, and assuming chemical equilibrium for both ions. We also include the depletion fraction of metals on dust as a function of gas-phase metallicity from \citet{decia2016} and \citet{Konstantopoulou2022} and apply the correction formula cell by cell in the simulation.

The ground-state energy level of the \cII\ and \siII\ ions is split into two spin states having an energy difference of 0.0079 eV for \cII\ and 0.0356 eV for \siII. These fine-structure levels produce two additional channels (referred to as fluorescent channels, noted as \cII* and \siII*) of photon absorption and emission at 1335.66~\AA\ and 1335.71~\AA, respectively, for \cII, and at 1264.73~\AA\ and 1265.02~\AA, respectively, for \siII. At typical ISM temperatures, nearly all of the C and Si will reside in the ground state. Under these conditions, large radiative excitation through the 1260 and 1334\AA\ channels leads to radiative de-excitation through either the resonant or the fluorescent channel. Since the densities of photons and particles are low, the fine structure state will radiatively decay back down to the ground state. In other words, this means that while the ground state might be optically thick to radiation, the fluorescent line can remain optically thin to radiative excitation such that fluorescent photons are not radiatively trapped as easily as in the resonant channel.
 
 Similarly to \citet{mauerhofer2021} and \citet{gazagnes2023}, we consider both the resonant and fluorescent channels of excitation and de-excitation when generating the LIS mock spectra. The final simulated line spectra are created using the radiative transfer post-processing code \textsc{rascas} \citep{rascas2009} using the following procedure: a total of one million photon packets sample the spectra of stellar populations contained in the star particles, within a wavelength range of $\pm$ 10 \AA\ around the \cIIl\ and \siIIl\ lines. Photon packets are emitted from star particles and travel through the simulation grid. In each cell the optical depth is the sum of all channels of interaction:

 \begin{equation}
 \begin{split}
     \tau_{\rm cell} = \tau_{\rm C_{II}\ 1334.53}\ +\ \tau_{\rm C_{II}*\ 1335.66}\ +\ \tau_{\rm C_{II}*\ 1335.71}\ +\\ \tau_{\rm Si_{II}\ 1260.42} + \tau_{\rm Si_{II}*\ 1264.73} + \tau_{\rm Si_{II}*\ 1265.02} + \tau_{\rm dust}
     \end{split}
 \end{equation}

 The optical depth of each line is the product of the ion column density in each cell and the ion cross section which is defined by Eq~(2) in \citet{mauerhofer2021}. The line parameters are taken from the NIST atomic database \citep{NIST}\footnote{\url{https://www.nist.gov/pml/atomic-spectra-database}}.
 We compute the turbulent velocity of the gas ($v_{\rm turb}$) accounting for the gas density and velocity in the neighboring cells, as in the star-formation recipe of e.g., \citet{trebitsch2021}. $\tau_{\rm dust}$ is set following the implementation of \citet{katz2022} where the dust-to-metal ratio is a double power-law function of the metallicity and is set to 0 in cells with temperatures above 10$^5$ K \citep[inspired from the results of][]{remyruyer2014}. 
 
We implement the radiative transfer post-processing method within a spherical volume with a radius equivalent to three times the galaxy's virial radius (R$_{\rm vir}$). Within this volume, we generate a set of 22,500 mock spectra for \cII\ and \siII\ lines. These 22,500 mock spectra correspond to 300 distinct lines of sight of observations applied to each of the 75 simulation outputs. The 300 directions, obtained with HealPix, are uniformly sampled across the sphere and are the same for each simulation output. We employ an aperture size of 1" diameter ($\sim$7 kpc at $z = 4.2$ and 7.9 kpc at $z = 3$).

\subsection{Observations}
\label{sec:dataobs}

In this work, we compare the mock spectra described in Section~\ref{sec:datasim} to observations from the \vandels\ survey \citep{McLure2018_vandels, Pentericci2018_vandels, Garilli2021_vandels}. The \vandels\ survey was conducted using the Visible Multi-Object Spectrograph (VIMOS) on the ESO Very Large Telescope (VLT), featuring a median spectral resolution of $R=580$. All \vandels\ observations were taken with the MR grism+GG475 order sorting filter with a 1" slit width and a minimum slit length of 7". This extensive survey comprises deep spectra of 2087 galaxies, covering observed wavelengths ranging from 4,800\AA\ to 10,000\AA. The survey encompasses regions in the Chandra Deep Field South (CDFS) and UKIDSS Ultra Deep Survey (UDS) fields. Notably, the majority of objects included in the \vandels\ dataset consist of star-forming galaxies within the redshift range of 2.4 $\leq z \leq$  7.0 such that VIMOS provides coverage of their rest-frame far-ultraviolet (FUV) spectra.

For this work, we focus on the analysis of the \cIIl\ and \siIIl\ absorption lines (including the nearby \cII* and \siII* fluorescent features). We pre-select a first sample of 965 galaxies at $z > 2.8$  to ensure that both the  \siII\ and \cII\ features are covered. We note that all the selected galaxies have high-quality spectroscopic redshifts \citep[\vandels\ quality flags of 3, 4, or 9 which means the redshifts are reliable at $>95\%$, ][]{Garilli2021_vandels}. Nevertheless, in some cases, the redshift estimations are based on either Lyman$-\alpha$ or ISM lines so the exact systemic redshift might differ by a few hundred km s$^{-1}$ (less or equal to the spectroscopic resolution element). We further detailed how we accounted for possible line shifts in Section~\ref{sec:fitting} where we directly compare the \vandels\ and mock spectra line profiles.

\begin{figure}[!htbp]
    \centering
        \includegraphics[width = \hsize]{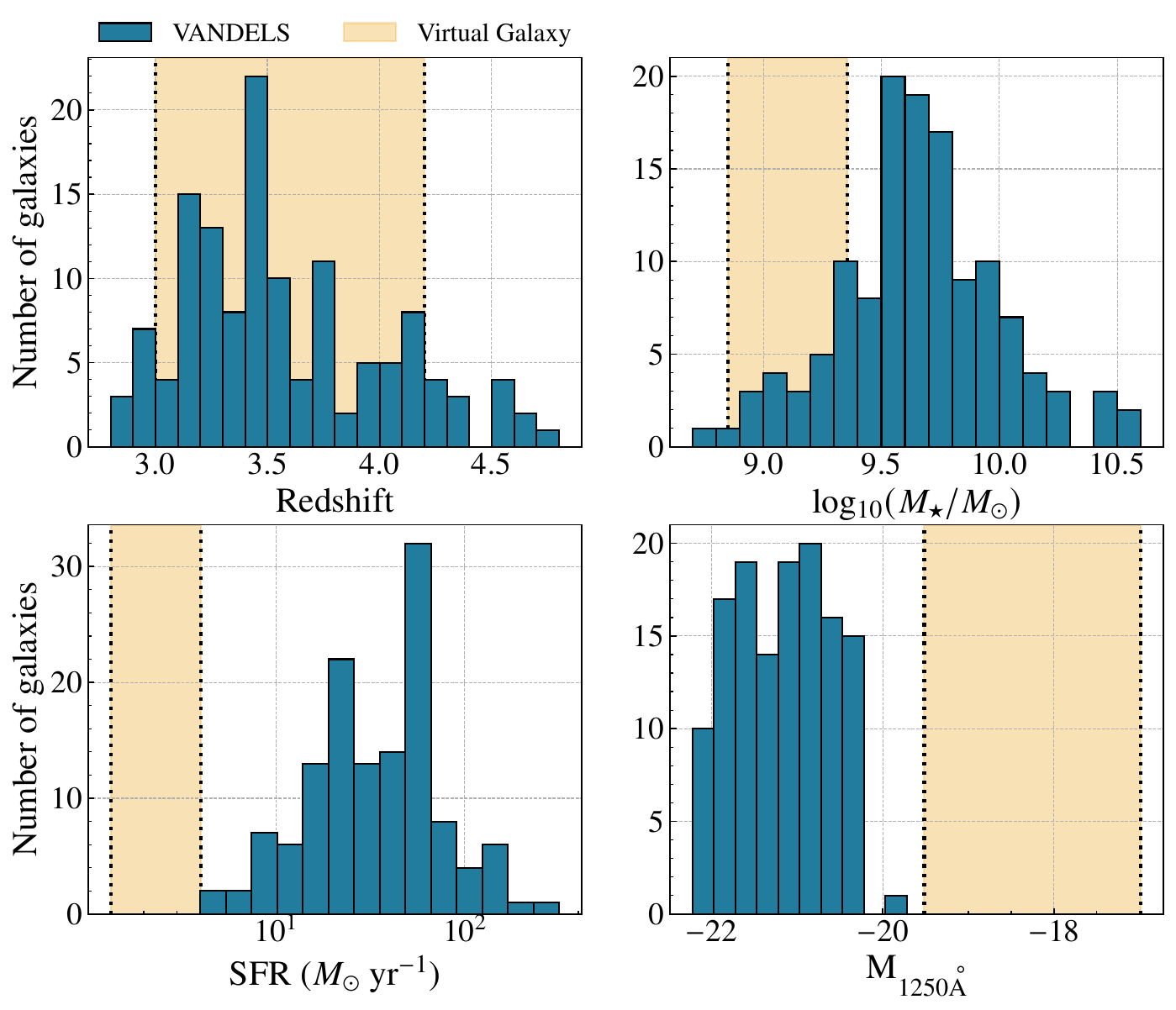}
    \caption{Histograms comparing the range of redshift, stellar mass, SFR, and absolute AB magnitude at 1250\AA\ for the 131 \vandels\ galaxies selected in this work ($z>2.8$ and S/N $>5$, in blue) and for the virtual galaxy (yellow). For the latter, the ranges shown are determined by the minimum and maximum values for each property over the 690 Myrs period used to generate the mock spectra.}
    \label{fig:orig_extract}
\end{figure}

From this first sample of 965 galaxies with both \siII\ and \cII\ observed, we select spectra that have a signal-to-noise ratio (S/N) of at least 5 per resolution element. We chose a high S/N cut because \vandels\ spectra have relatively low resolution, so the number of spectral bins that can be directly compared between the simulated and observed spectra is small. Selecting high S/N spectra ensures that observations have reduced noise features, reducing the chance of biasing our comparisons. We visually checked each \vandels\ spectrum to ensure there was no interloper contamination around the \cII\ and \siII\ lines. We ended up with a final set of 131 galaxies.  Figure~\ref{fig:orig_extract} illustrates the redshift, $M_\star$, SFR, and absolute AB magnitude at 1250\AA\ (M$_{1250\AA}$) ranges of our final \vandels\ sub-sample alongside those of the virtual galaxy, considering its evolution from $z = 4.19$ to $z = 3.00$. In \citet{gazagnes2023}, we compared the current simulation to a sample of $z<0.18$ galaxies with a broad range of stellar mass ($\sim 10^6 - 10^{10}$~$M_\odot$). In contrast, the \vandels\ sample better matches the redshift range of the virtual object, and its range of M$_\star$ is slightly larger than that of our virtual galaxy ($\sim 10^9 - 10^{10.5}$~M$_\odot$).  

The SFR and  $M_{1250\AA}$ of the \vandels\ galaxies are significantly larger than our virtual galaxy. Yet, in \citet{gazagnes2023}, we showed that the simulated spectra, derived from the same virtual galaxy, successfully replicated observations from galaxies within the mass range of  $\sim10^6 - 10^{9.5}M_\odot$ and SFRs between 10$^{-2}$ to 10$^{1.58}M_\odot$ yr$^{-1}$. Hence, the lack of an exact match between the properties of the virtual galaxy and observed counterparts does not present a significant obstacle to comparing the LIS line spectra. We discuss further this aspect in Section~\ref{sec:galprop}. In general, the selected \vandels\ galaxies, which cover substantially larger $z$, M$_\star$, SFR, and magnitudes as compared to \classy, is an excellent sample to test the conclusions from our previous study.

\section{Comparing the \cII\ and \siII\ spectra in VANDELS and in the simulation}
\label{sec:measurement}
Here, we compare the simulated \cIIl+\cIIlstar\ and \siIIl+\siIIlstar\ spectra and the \vandels\ observations. We first explore the overlap in the lines' equivalent widths in Section~\ref{sec:linesmeasurements}, and then identify the mock spectra that best resemble the observed line profiles of each \vandels\ galaxy in Section~\ref{sec:fitting}.

\subsection{Overview of \cII\ and \siII\ measurements}
\label{sec:linesmeasurements}
In this section, we measure the EWs of the \cIIl\ and \siIIl\ spectra to compare the overall consistency of the lines' properties in \vandels\ and in the virtual galaxy. The spectral resolution of the \vandels\ spectra is relatively low, causing the \cIIl\ and \cIIlstar\ features (separated by $\sim$1.2~\AA, equivalent to half a resolution element) to blend. For comparison, the separation of the \siIIl\ and \siIIlstar\ features is approximately 4.5~\AA\ apart, equivalent to roughly two resolution elements. To simplify the comparison of the \cII\ and \siII\ line spectra, we choose to measure the combined EW(\cIIl\ + \cIIlstar) and EW(\siIIl\ + \siIIlstar).

The measurement method is as follows: we normalize the \siII\  and \cII\ simulated spectra using the median of the flux using a feature-free continuum interval around 1255 \AA\ and 1331 \AA, respectively and convolve them to the resolution of the \vandels\ spectra (R$\sim$580).  We derive the final EW measurements by integrating the continuum-normalized flux in an interval of $\pm$ 2000~km~s$^{-1}$ around $\lambda = 1334.53$~\AA\ and $\lambda = 1260.42$~\AA, respectively. Spectra dominated by absorption (emission) have positive (negative) EW values. We note that there exists a \ion{S}{2} absorption line at 1259.52~\AA, blended with \siIIl, which is not included in our radiative transfer model. In \citet{gazagnes2023}, we showed that its contribution, resolved in higher-resolution spectra, is typically much weaker than \siIIl\ and has a negligible impact on the line measurements.

\begin{figure}[!htbp]
    \centering
    \includegraphics[width = \hsize]{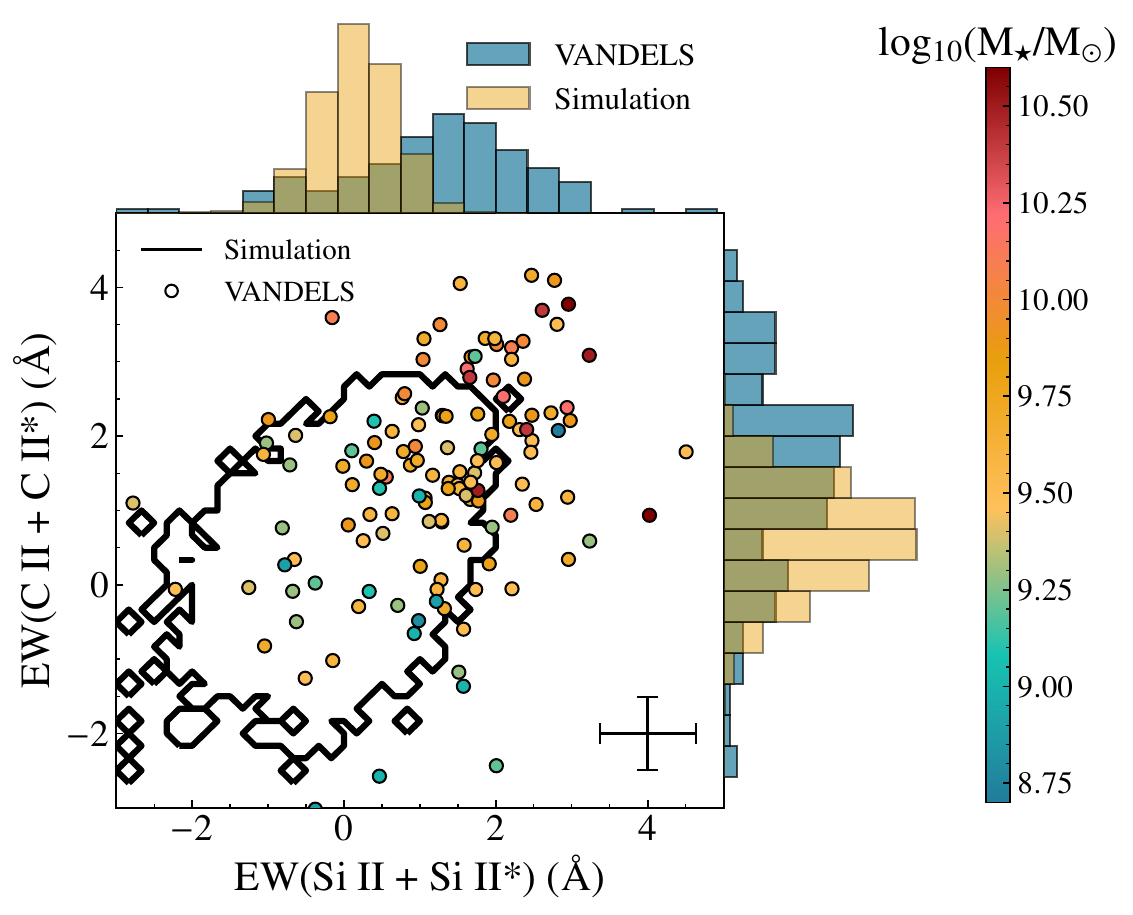}
    \caption{The comparison of the EW(\cIIl\ + \cIIlstar) and EW(\siIIl\ + \siIIlstar) in the simulation (black contour) and in the \vandels\ observations (circles). The \vandels\ points are color-coded by their stellar mass (the virtual galaxy has $M_\star$ within $\sim 10^9 - 10^{10.5}$~$M_\odot$), and we show in the bottom right the typical uncertainty on the measurements. The black contour includes the entire range of the mock spectra measurements. The \vandels\ galaxy sample exhibits a marginally wider range of EW values, featuring notably larger values in comparison to the simulation. }
    \label{fig:glob}
\end{figure}

Figure~\ref{fig:glob} compares the EW(\cIIl\ + \cIIlstar) and EW(\siIIl\ + \siIIlstar) values in both the simulation and the \vandels\ observations. The \cII\ EW values scale strongly with the \siII\ EWs, highlighting that the properties of both ions are closely related. Further, the correlation and scatter in the EW values of the simulation align well with observational data. We observe a slight shift towards higher EWs (above 2.5\AA) in the observations, with 33\% of the \vandels\ spectra measurements falling outside the range of simulated values. This difference is likely attributable to the generally larger stellar masses of the \vandels\ galaxies compared to our simulated galaxy, and these objects typically have broad absorption profiles ($>1000$ km s$^{-1}$ in width) unseen in the virtual galaxy \citep{gazagnes2023}. 

The scatter in the simulation's EW values arises because we look at several independent lines of sight, each revealing diverse ISM conditions. This results in a fluctuation of EW values around typical peaks for both ions, as seen in the 1D distribution. Similarly, this line-of-sight variability might also contribute to the scatter observed in real-world data, as each observation captures a unique perspective of the targeted galaxy \citep[see also the discussion in the context of comparing the simulation to the \classy\ observations, ][]{gazagnes2023}.

In the following section, we dive deeper into the comparison of the simulated and observed spectra and analyze the similitude of the line spectral shapes in both datasets.

\subsection{Finding the best-matching mock spectra}
\label{sec:fitting}

\begin{figure*}[!htbp]

    \centering
        \includegraphics[width =0.95 \hsize]{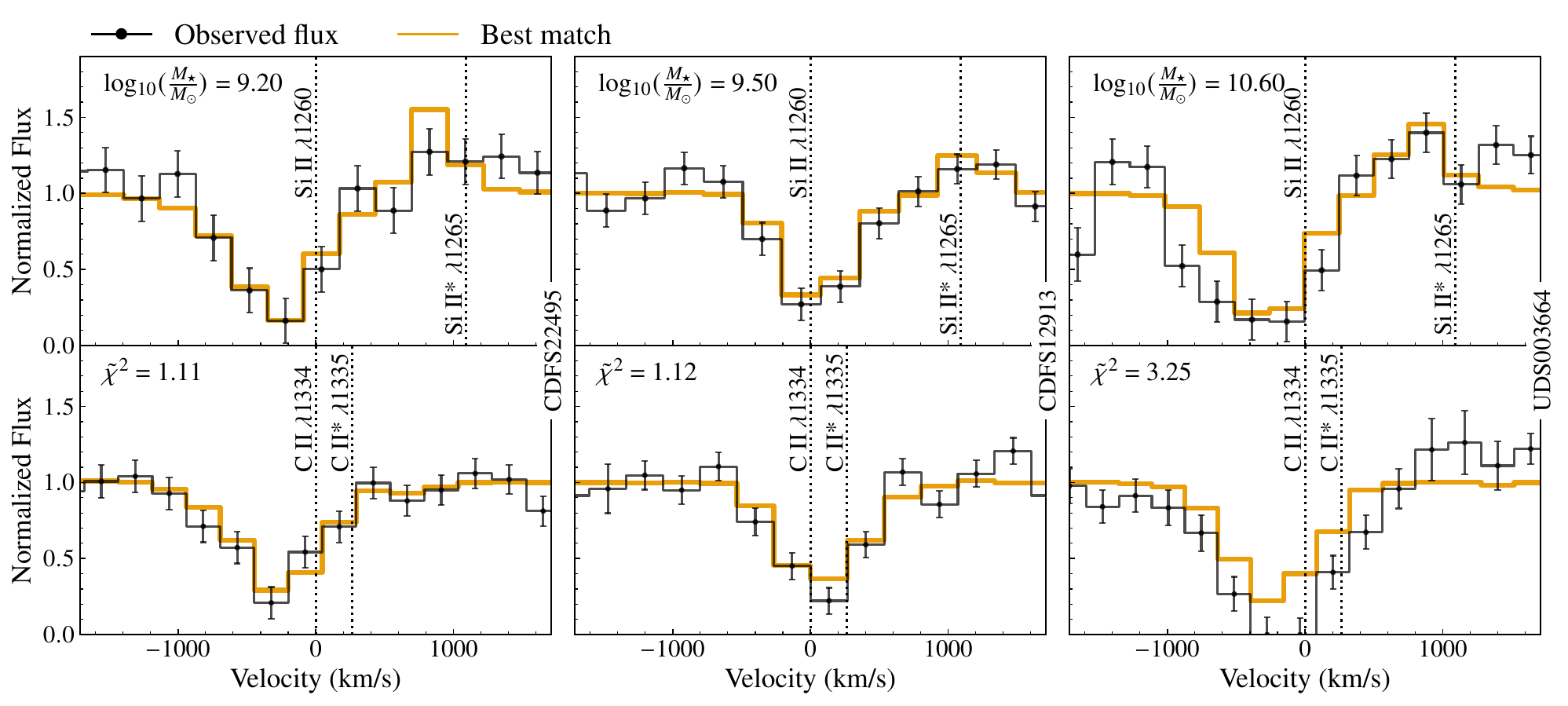}
    \caption{The best-matching spectra of the \siIIl+\siIIlstar\ (top panels) and \cIIl+\cIIlstar\ (bottom panels)  line profiles for three \vandels\ galaxies of different M$_\star$ (shown the top left corner of the top panels) using the procedure detailed in Section~\ref{sec:fitting}. The black and orange spectra are the observed flux and best-matching spectrum (lowest $\tilde{\chi}^2$ values, indicated in the top left corner of the bottom panels), respectively. The vertical dotted lines indicate the expected locations of each transition. The higher mass galaxy, which presents broad absorption profiles, is not well matched by the lowest $\tilde{\chi}^2$ spectrum.  }
    \label{fig:fitexamples}
\end{figure*}

Comparing the EW(\cIIl\ + \cIIlstar) and EW(\siIIl\ + \siIIlstar) only offers limited insights into the resemblance of the line profiles. This is because we combine the properties of both resonant and fluorescent features, overlooking the specific shape of the lines. A direct comparison of the shape of UV absorption line profiles is crucial for gaining insights into the geometry of the absorbing gas \citep[see e.g.,][]{scarlata2015,carr2018, carr2022}, in particular for comparing how distinct models reproduce the data \citep[e.g clumpy CGN versus homogeneous wind model,][]{li2024_alpaca}. In this section, we directly assess the spectral similitude between the simulated and \vandels\ observations.

To achieve this, we look for the best-matching mock spectra from the simulation for each \vandels\ spectrum. We employ the same minimization procedure as outlined in \citet{gazagnes2023}. The mock and observed profiles are normalized using the median flux within feature-free intervals located around 1256 \AA\ and 1331 \AA\ for the \siII\ and \cII\ spectra, respectively. Then, the 22,500 mock spectra are convolved to the resolution of the \vandels\ observations (R$\sim$580). For each mock spectrum, we compute the reduced $\tilde{\chi}^2$ value:

\begin{equation}
\label{eq:chi2}
\begin{split}
    \tilde{\chi}^2 &= \frac{\chi^2_{\rm C\ II} +\chi^2_{\rm Si\ II}}{n_{\rm C\ II}^{\rm obs} + n_{\rm Si\ II}^{\rm obs}} , \text{ where} \\
    \chi^2_{\rm C\ II} &= \sum{\left( \frac{F_{\rm C\ II}^{\rm obs} -  F_{\rm C\ II}^{\rm sim}}{\sigma_{\rm C\ II}^{\rm obs}} \right)^2} \text{ and }  \\  
    \chi^2_{\rm Si\ II} &=\sum{\left(\frac{F_{\rm Si\ II}^{\rm obs} -  F_{\rm Si\ II}^{\rm sim}}{\sigma_{\rm Si\ II}^{\rm obs}} \right)^2}.
\end{split}
\end{equation}

\noindent $F_{\rm ion}^{\rm obs}$ and $F_{\rm ion}^{\rm sim}$ are the observed and simulated flux in the wavelength regions of each ion, respectively. $\sigma_{\rm ion}^{\rm obs}$ represents the error on the observed flux, and $n_{\rm ion}^{\rm obs}$ is the number of wavelength bins used to compare both spectra.


In our analysis, we address the redshift uncertainty (\vandels\ redshift flag of 3, 4, or, 9, meaning their redshift is reliable at $>95\%$, see Section~\ref{sec:dataobs}) in our fits by initially conducting a cross-correlation between the observed spectra and each mock spectrum, allowing for the observations to be shifted by at maximum one spectral resolution element. This cross-correlation process determines the optimal shift required to align the spectral features. However, we acknowledge that this approach tends to cancel out any actual differences in the line velocity offsets between the mock and observed spectra. To assess the potential impact of this pre-alignment step, we conducted an independent experiment wherein we identified the best-matching mock spectrum both with and without including the cross-correlation step. While we observed that the typical $\tilde{\chi}^2$ of the best match was larger and some best-matching spectra did not align as well when excluding the cross-correlation step, we find that this does not have a significant impact on the overall results and subsequent discussions presented in the following sections. We elaborate on the outcomes of these tests in Appendix~\ref{app:offseteffect}, highlighting that employing an alternative strategy yields similar results.

In \citet{gazagnes2023}, we selected a unique, best-matching mock spectrum for each observed galaxy. Here, for each \vandels\ object, we also look at the distribution of the $\tilde{\chi}^2$ for all 22,500 mock spectra to explore the diversity of spectral shape from the simulation that closely mimics a given observation. This approach enables us to further assess the simulation's accuracy in replicating a specific observation by evaluating the number of mock spectra that closely align with it. It also facilitates the examination of common features or characteristics among all mock spectra that well match with the observation, an aspect that we discuss in detail in Section~\ref{sec:lycpre}.

In Figure~\ref{fig:fitexamples}, we present the matching outcomes for three galaxies drawn from our selected \vandels\ sample. We chose representative galaxies covering/spanning a range of stellar masses, specifically $10^{9.2}$, $10^{9.6}$, and $10^{10.6}$ ${\rm M_\odot}$. For each galaxy, we show the best-matching mock spectrum (i.e., with the lowest $\tilde{\chi}^2$ value). The best-matching spectrum replicates reasonably well the line profiles in the two lower-mass galaxies ($\tilde{\chi}^2 < 1.5$). However, they fall short of capturing the complete width of the broad absorption lines observed in the $10^{10.6}$ ${\rm M_\odot}$ galaxy ($\tilde{\chi}^2 > 3$). This result reaffirms the findings from \citet{gazagnes2023, blaizot2023simulating}, emphasizing that mock line spectra generated from a $10^{9.2}$ ${\rm M_\odot}$ virtual galaxy do not provide an accurate match for the most massive galaxies, where the ISM gas environment may be significantly different (e.g., more turbulent or larger outflows). 

\begin{figure}[!htbp]
    \centering
        \includegraphics[width =\hsize]{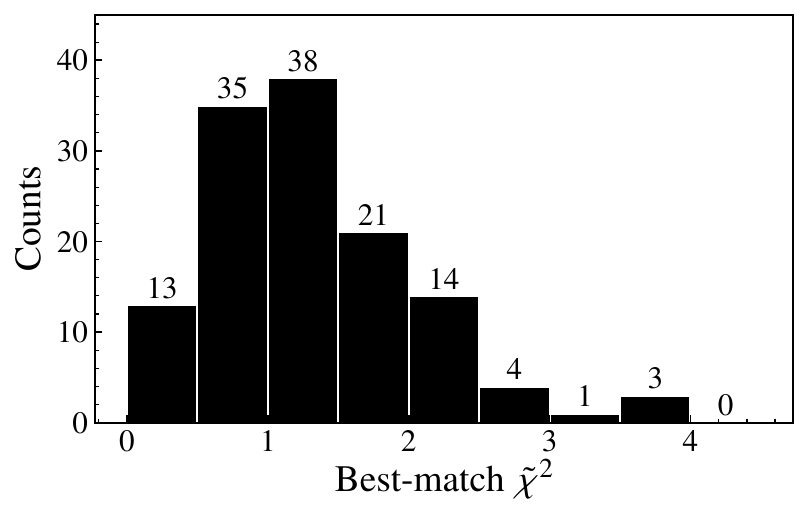}
    \caption{Histogram of the best-match $\tilde{\chi}^2$ for the 131 \vandels\ spectra analyzed in this work. 37\% of the spectra are well reproduced ($\tilde{\chi}^2 < 1$), and 83\% are reasonably matched ($ \tilde{\chi}^2<2$).}
    \label{fig:chisquare}
\end{figure}

Figure~\ref{fig:chisquare} shows the overall distribution of $ \tilde{\chi}^2$ for the whole sample. The top panel presents the distribution of the $\tilde{\chi}^2$ value of the best-matching spectrum for each object. 37\% of the spectra are well reproduced ($\tilde{\chi}^2 < 1$), and 83\% reasonably matched ($ \tilde{\chi}^2<2$). These numbers are slightly lower than for the \classy\ sample of $z\sim0$ galaxies (60\% with $\tilde{\chi}^2 < 1$, 89\% with $\tilde{\chi}^2 < 2$). These results may be explained by the presence of more massive galaxies in the selected \vandels\ sample whose spectra are more complex to match using the current simulation. We discuss further the influence of galaxy properties such as $M_\star$ and SFR on the agreement between the best matches and observations in the next section.

\section{Connecting the best-matched spectra with galaxy properties}
\label{sec:galprop}

Section~\ref{sec:fitting} showed that mock spectra, derived from a single virtual galaxy, reasonably well replicate ($\tilde{\chi}^2 < 2$) 83\% of the selected \vandels\ galaxies exhibiting a diverse range of line profiles and characteristics. These results echo those previously discussed in the context of the comparison with the low-$z$ \classy\ sample in \citet{gazagnes2023}. In that study, we emphasized that monitoring the virtual galaxy's evolution over 690 million years and exploring it from various viewing angles enabled us to encounter a wide range of ISM configurations characterized by evolving properties. In this section, we further explore the relationship between the best matches and galaxy properties, examining which factors determine the agreement between simulated and observed line spectra. Section~\ref{sec:compsfr} explores the influence of M$_\star$ and SFR for finding the best match to a given observation, and Section~\ref{sec:spec_ori} delves deeper into the role of star formation bursts into producing LIS spectral features that resemble the observations.

\subsection{The influence of $M_\star$ and SFR when matching simulations and observations}
 \label{sec:compsfr}

\begin{figure}[!htbp]
    \centering
        \includegraphics[width =\hsize]{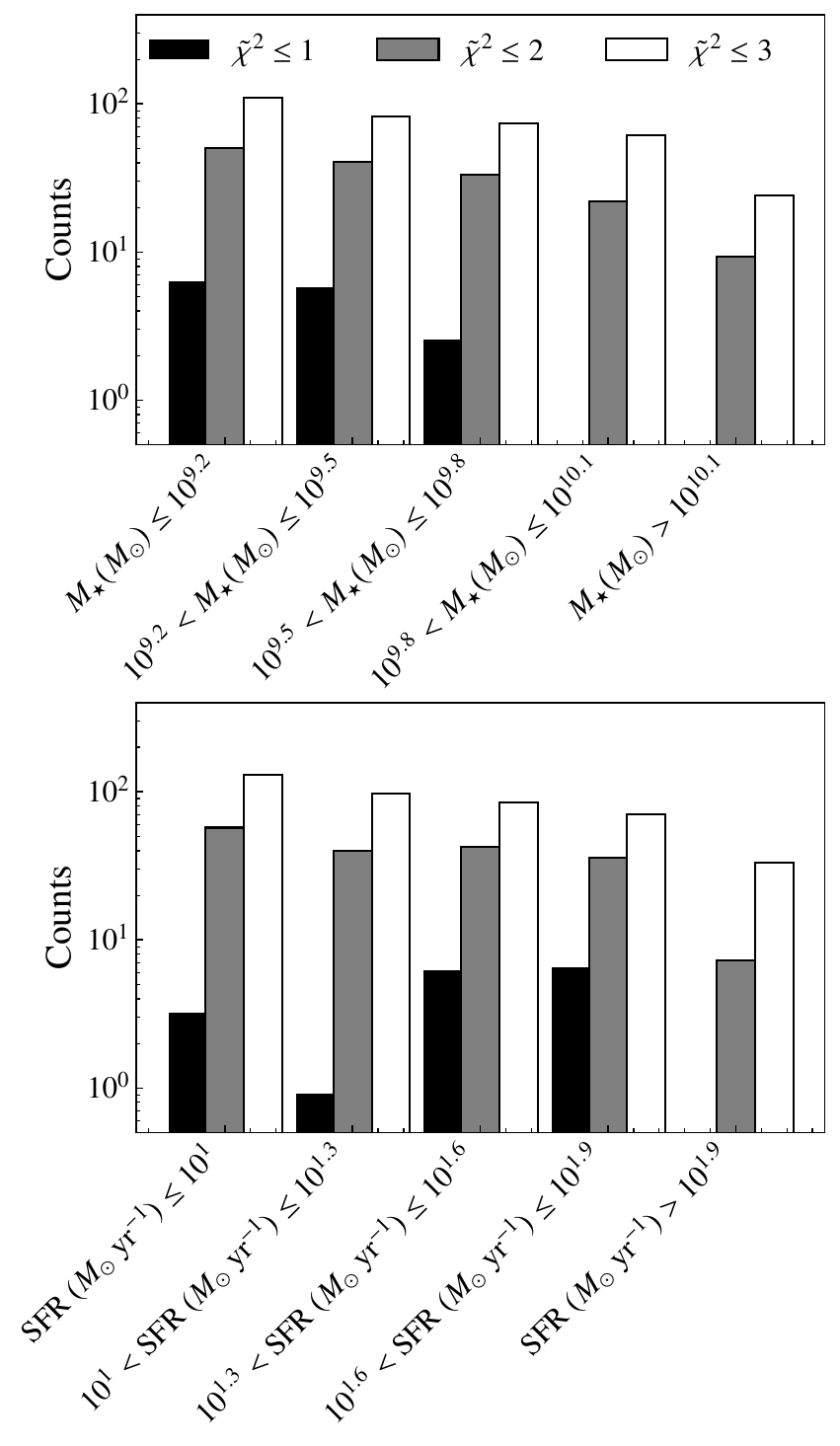}
    \caption{Top: histogram presenting the average number of mock spectra with $\tilde{\chi}^2 < 1$ (black), $\tilde{\chi}^2 < 2$ (gray), and $\tilde{\chi}^2 < 3$ (white) per observed galaxy, in bins of stellar mass. The most massive \vandels\ galaxies ($\sim > 10^{10}$) are not well replicated by the mock spectra for our virtual galaxy. Bottom: Same but for the SFR. }
    \label{fig:chisquare_Mstar}
\end{figure}

In this section, we discuss the influence of $M_\star$ and SFR on the matching quality. Due to the lack of precise overlap between $M_\star$ and SFR between the simulation and \vandels\ (see Fig~\ref{fig:orig_extract}), we investigate whether a greater number of mock spectra closely match a given observation when the corresponding \vandels\ galaxy has a stellar mass and SFR more similar to those of the virtual galaxy. To do so, we analyze the distribution of the average number of mock spectra with $\tilde{\chi}^2$ values less than 1, 2, and 3, respectively, per observed galaxy, across bins of increasing mass and SFR.

The top panel of Figure~\ref{fig:chisquare_Mstar} shows the distribution of the average number of mock spectra with $\tilde{\chi}^2$ values less than 1, 2, and 3 per galaxy across increasing mass bins. The first bin ($M_\star < 10^{9.3}M_\odot$) directly aligns with the stellar mass range of the virtual galaxy ($\sim10^{8.9}$ to $10^{9.3} M_\odot$). Interestingly, this bin exhibits the highest average number of mock spectra with $\tilde{\chi}^2$ values less than 1, 2, and 3, indicating a greater resemblance between these mock spectra from the simulation and the observations within the same mass range.

Specifically, for galaxies with $M_\star < 10^{9.8}M_\odot$, there are, on average, 5 to 8 mock spectra with $\tilde{\chi}^2 < 1$, while none are found for galaxies in the two largest stellar mass bins ($\sim > 10^{10}M_\odot$). These results echo \citet{gazagnes2023, blaizot2023simulating}, indicating that the current simulation may not accurately represent galaxies with $M_\star > 10^{10}M_\odot$. Furthermore, we observe a declining trend in the average number of mock spectra with $\tilde{\chi}^2 < 2$ and $\tilde{\chi}^2 < 3$ as a function of stellar mass. This trend suggests that as galaxies increase in mass, it becomes increasingly challenging to find multiple mock spectra closely resembling the observed data.

The bottom panel of Figure~\ref{fig:chisquare_Mstar} focuses on the SFR. We note that in \vandels, no galaxy exhibits an SFR consistent with the SFR of our virtual galaxy ($\sim10^{0}$ to $10^{0.6} M_\odot$ yr$^{-1}$). This panel shows that, similarly to the analysis for stellar mass, the bins representing the most extreme star-forming galaxies yield no matches with $\tilde{\chi}^2 < 1$. Given that these galaxies are also the most massive, this outcome aligns with the observations from the top panel. We do not observe a declining trend in the average number of mock spectra with $\tilde{\chi}^2 < 1$ from the lowest SFR bin to the largest SFR bin. However, we do discern a slight decrease in the average number of $\tilde{\chi}^2 < 2$ and $\tilde{\chi}^2 < 3$, suggesting that, similarly to $M_\star$, albeit to a lesser extent, it becomes increasingly challenging to find a spectrum that adequately reproduces an observation as the galaxy exhibits a higher SFR.

In general, our results reveal the existence of certain dependencies on both M$_\star$ and SFR. Specifically, the spectra of the most massive or star-forming galaxies in \vandels\ are not well matched in the simulation. Conversely, a greater number of mock spectra resemble observations with properties similar to those of the virtual galaxies. However, we also find that an exact match between observation and simulation properties is not strictly necessary to find mock spectra resembling real galaxies. This outcome likely stems from the extensive sampling of 300 viewing angles over a 690 Myr period, which unveils rarer ISM and CGM configurations that may be more prevalent in galaxies with distinct properties.

Previous studies revealed noticeable yet scattered dependencies of LIS absorption line properties on M$\star$ and SFR \citep[e.g., the EW and velocity offsets;][]{rupke2005_outflow,chisholm2015_scaling, chisholm2016_photoionizingoutflow}.  Our findings may provide important clues to understanding the scatter observed in typical line property to stellar mass trends in those studies. Given that galaxies are not idealized symmetric objects, there exists a distribution of ISM and CGM configurations observable depending on the viewing angle. The peak and deviation of this distribution depend on galaxy properties, but our measurements only capture a unique point in this distribution contingent upon our line of sight which may deviate from this peak.

Overall, Fig~\ref{fig:chisquare_Mstar} suggests that the fits we obtain are genuine rather than coincidental, i.e. that the scatter in the mock spectra reflects physical variations in time and line-of-sight that are bounded in a range loosely set by the mass of the system. Achieving good agreement between observed and simulated spectra doesn't necessarily imply an exact match between the virtual galaxy environment and the object considered, but it suggests that the observed galaxy falls within a certain dynamical range that can be replicated with a given simulation. In turn, analyzing the most massive objects requires the inclusion of virtual galaxies with properties closer to the observed range \citep[e.g. using those from the \textsc{sphinx} simulations]{rosdahl2018}. While this is an important effort for delineating the range of spectral properties that single-galaxy simulations can accurately reproduce as a function of their properties, this effort extends beyond the scope of the present paper.

\subsection{Best-matching spectra originate from luminosity peaks close to star-formation bursts}
\label{sec:spec_ori}

\begin{figure}[!htbp]
    \centering
        \includegraphics[width =\hsize]{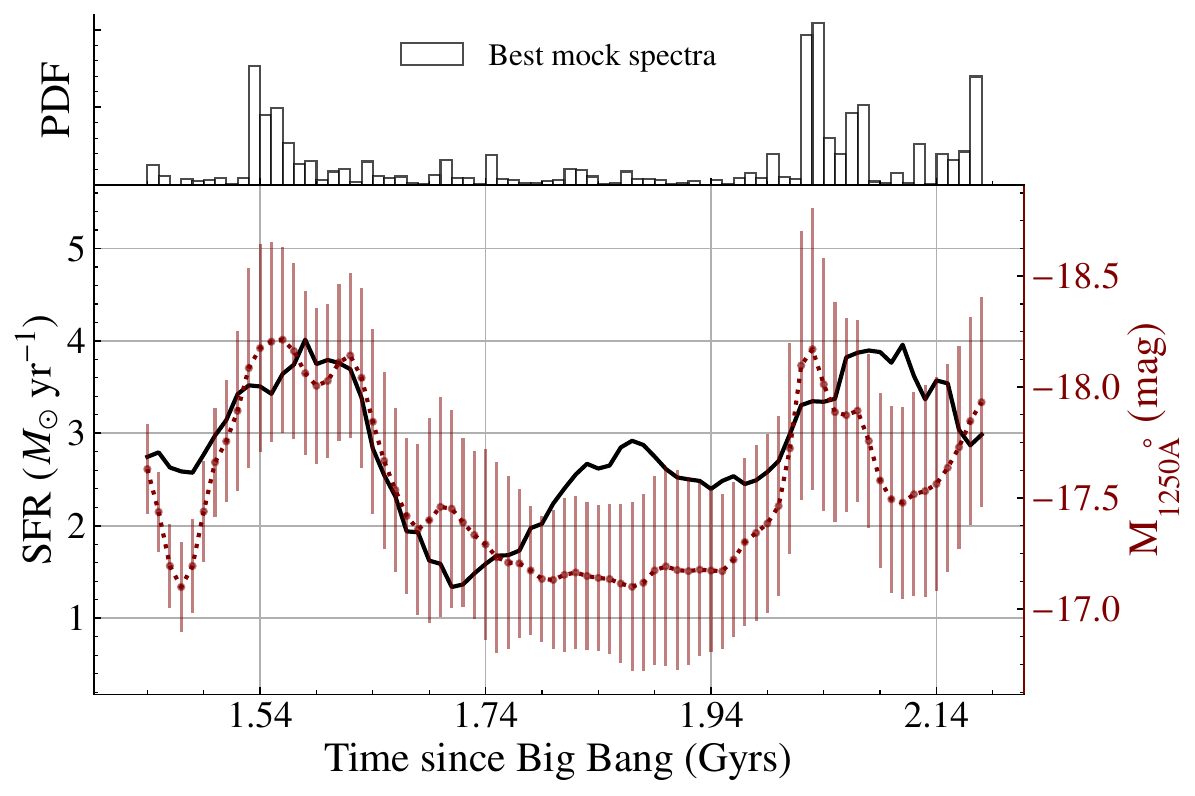}
    \caption{The evolution of the SFR (black) and AB magnitude at 1250~\AA\ (red) of the virtual galaxy over the 75 time steps used to produce the mock \cII\ and \siII\ line profiles (690 Myrs period). The magnitude curve is derived using the median and standard deviation from the 300 line-of-sight measurements at each time step. At the top, we present a histogram illustrating the distribution of time steps where the 2,790 best-matching mock spectra are found (taking 30 best mocks for each of the 93 \vandels\ galaxies selected in Section~\ref{sec:fitting}). Overall, the best-matching spectra dominantly originate from luminosity peaks closely preceding the two most intense star-formation bursts.}
    \label{fig:origin_sfr}
\end{figure}



An important and novel aspect of our analysis involves examining the average distribution of $\tilde{\chi}^2$ values across multiple mock spectra for each galaxy. This strategy allows us to establish a more robust understanding of the range of mock spectra that closely match the observed spectrum, as opposed to relying solely on a single best-matching spectrum. In the subsequent sections, we build upon this distribution to interpret our results, focusing on a smaller subset of \vandels\ galaxies that have been reasonably well replicated by the simulation. Specifically, we consider the 93 \vandels\ galaxies with at least 30 mock spectra with $\tilde{\chi}^2 < 2$ as suitable candidates for further investigating the origin of the good agreement between simulation and observation. This selection criterion is derived from the insights presented in the lower panel of Figure~\ref{fig:chisquare}, which demonstrates that galaxies within the two highest stellar mass bins fail to meet this threshold.

In this section, we explore further the connection between the virtual galaxy star-formation history and the origin of the good agreement between observations and simulations. To do so, we look at the distribution of time steps at which each of the best mock spectra was generated and investigate whether they originate from specific stages in the galaxy's evolutionary timeline. To do so, we extract the corresponding 2,790 best-matching mock spectra (30 mock spectra $\times$ 93 well-matched \vandels\ galaxies).


In Figure~\ref{fig:origin_sfr}, we plot the evolution of the SFR of the virtual galaxy over the period considered in this work. We measure the AB magnitude of the virtual galaxy taking the flux at 1250 \AA\ (M$_{1250\text{\AA}}$) for each output and direction and plot the evolution of the median and standard deviation values from 300 directions. The distribution of the best-matching spectra is very tightly correlated with the evolution of M$_{1250\text{\AA}}$, and most of the best mocks originate from bright luminosity peaks. Interestingly, these peaks typically precede strong star-formation bursts where the galaxy is at its peak of star formation. However, we note that the SFR of the virtual galaxy at time $t$ is calculated as an average over the preceding 100 Myrs. As a result, the luminosity peaks may coincide with SFR peaks, albeit with a delay in its influence on the 100-Myrs averaged curve.

We find that 85 out of 93 selected \vandels\ galaxies exhibit at least one best-matching mock spectrum originating from either the 1.53 Gyrs or the 2.02 Gyrs time steps, corresponding to the two highest peaks in the histogram distribution near the peaks of SFR. This finding has two implications. Firstly, it indicates that a comparable success rate in matching spectra to observations could have been achieved by solely considering these two time steps. Secondly, given the diversity of line properties exhibited by the 93 \vandels\ galaxies (shown in Fig~\ref{fig:glob}), it implies that the LIS profiles generated at these times encompass a wide range of variability such that two distinct viewing angles likely have significantly different line profiles. We note again that, while a strong agreement between observed and simulated spectra during these two bursts is notable, it does not systematically imply an exact match between the environments of the observed and simulated objects.

The fact that the best-matching mock spectra aren't randomly distributed in time but rather linked to specific star-formation phases is a fundamental result as it suggests that these phases have a direct impact on the LIS metal line spectra. We highlighted in \citet{gazagnes2023} that one of the main differences in the line properties was that the absorption lines are significantly more redshifted during periods of relative quiescence as compared to star-formation events (see Figure~10 in particular). This suggests that the gas kinematics may be one powerful tracer of the current star-formation activity \citep[see also][]{Trebitsch2017, rosdahl2018}.   

As detailed in Section~\ref{sec:compsfr}, although the virtual galaxy properties do not precisely match those of the selected \vandels\ sample, we still identify mock spectra closely resembling the observations. This means that, in the context of understanding how LIS metal line spectra connect to the dominant physical processes, the relative evolution of the SFR of a galaxy is likely more relevant than its absolute fluctuations at a certain time. One can also note the four-fold variation in brightness, closely tied to the SFR, of the virtual galaxy. Hence, at high $z$, where galaxies may exhibit bursty behavior due to ample neutral hydrogen reservoirs, the observation of bright luminous objects may be systematically biased toward objects experiencing intense star-formation bursts. However, these objects may not be as massive as their luminosity would imply since they are caught in a very specific phase of their evolution.


To conclude on a rather general note, Figure~\ref{fig:origin_sfr} has critical implications for understanding existing observational biases inherent in current high-$z$ surveys. \vandels\ is a UV continuum selected sample, meaning that it is implicitly biased towards UV-bright galaxies \citep{Garilli2021_vandels}  likely in a phase of star-formation (a bias that we further accentuated by selecting the highest SNR objects). Interestingly, when matching mock and observed spectra, we recover the selection function of the observed survey because the best mock spectra correspond to times when the virtual galaxy is also UV-bright. This means that the spectral properties of such galaxies are significantly influenced by their current stage in the evolutionary cycle, differing quite notably from those in less luminous or less active star-forming phases.

\section{Predicting the leakage of ionizing photons in \vandels}
\label{sec:lyc}

In the high-$z$ universe, the opacity of the IGM prevents any robust measurements of the LyC escape fraction from direct observations of the flux below 912\AA. Nevertheless, the study of low$-z$ LCEs has enabled us to determine indirect diagnostics to \fesclyc\ \citep[e.g., O$_{32}$, \lya\, \ion{Mg}{2}, $\beta$ slopes, or LIS covering fractions, see ][]{jaskot2013, verhamme2017, izotov2023_lyaemission_deficientmetal, chisholm2020, chisholm2022_beta, flury2022_lycdiag, gazagnes2018, reddy2016stack, saldana2023}. Although empirical trends between \fesclyc\ and these diagnostics show a certain scatter, they are already used to put the first constraints on the escape fractions of reionization-era star-forming galaxies \citep[e.g ][]{Choustikov2023_, Witten2023_lycjwst}. Here, for each \vandels\ galaxy in our sample, we predict \fesclyc\ using the simulation and compare these predictions with previous constraints from \citet{Begley2022_lyc}, \citet{Begley2023_lya_lyc}, and \citet{saldana2023}.  Section~\ref{sec:lycpre} details our simulation-based approach to predict \fesclyc\ for each \vandels\ galaxy, Section~\ref{sec:lycexp} explores the consistency of these predictions with other known LyC diagnostics, and Section~\ref{sec:lycsimu} investigates what makes LIS metal lines good tracers of the LyC leakage.

\subsection{Extracting simulation-based \fesclyc\ predictions}
\label{sec:lycpre}

\begin{figure}[!htbp]
    \centering
        \includegraphics[width =\hsize]{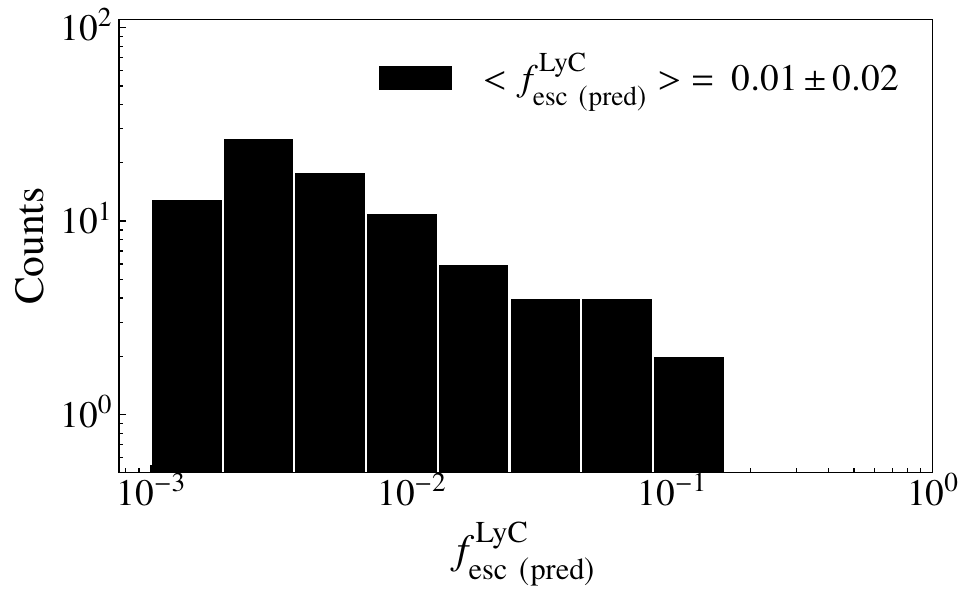}
    \caption{The distribution of  \fesclycpre for the \vandels\ sample analyzed in this work. The \fesclycpre\ is determined by taking the average \fesclyc\ value across the 30 best mock spectra for each of the 93 selected \vandels\ objects. The average \fesclycpre\ for this sample is 0.01 $\pm$ 0.02.}
    \label{fig:escapelyc}
\end{figure}

The RHD simulation used in this work includes a self-consistent propagation of the ionizing photons through radiative transfer, meaning that we can follow for each line of sight the amount of ionizing photons escaping the galaxy. To determine \fesclyc\ in the simulation, we take the approach of \citep{mauerhofer2021}: for each of the 300 directions and 75 outputs, we generate mock spectra of the virtual galaxy between 10\AA\ and 912\AA, such that each ionizing photon produced encounter an optical depth that is a sum of contributions from H, He, He+, and dust. We then compute the total number of ionizing photons going out of the virial radius and divide it by the intrinsic number produced to get \fesclyc. In this work, we only focus on the escape fraction at 900\AA, determined by averaging the spectra between 890\AA\ and 910\AA, to maintain consistency with LyC constraints in the existing literature. We note that for the current virtual galaxy, in the absence of hard ionizing sources, the escape fraction at 900\AA\ closely relates to the total escape fraction of ionizing photons \citep[see ][for more details on this aspect]{mauerhofer2021}.

Using the detailed strategy, we determine the LyC escape fraction corresponding to each of the 22,500 mock spectra generated. Then, to extract a \fesclyc\ prediction for each of the 93 \vandels\ best-matched objects, we take the median and standard deviation of the  30 \fesclyc\ values corresponding to the 30 best-matching mock spectra. To differentiate our predictions from direct LyC measurements, we refer to these predicted \fesclyc\ as \fesclycpre.


Figure~\ref{fig:escapelyc} presents the distribution of the 93 \fesclycpre\ measurements. The vast majority of values fall below 1\% escape, and only two objects have \fesclycpre\ $>$ 10\%. The average \fesclycpre\ for the 93 galaxies is 0.01 $\pm$ 0.02. While lower, this estimate falls within 2$\sigma$ of the 0.07$\pm$0.02 escape fraction reported by \citet{Begley2022_lyc} who directly constrained \fesclyc\ using ultra-deep, ground-based, U-band imaging. Interestingly, our average predicted escape fraction is highly consistent with \citet{saldana2023} who determined an average \fesclyc\ of 0.02$\pm$0.01 using an indirect approach combining the residual flux of LIS absorption lines (including \siIIl, 1302, 1527, \ion{O}{1}~$\lambda1302$, and \cIIl) and the expected dust attenuation at 912\AA\ \citep[Eq. (5) from][]{chisholm2018}. The consistency of both approaches is interesting, particularly because the predictions derived in this work do not require an explicit forecast of the dust attenuation at 912\AA. We further discuss this point in Section~\ref{sec:lycsimu}.


Using RHD simulations to predict the escape fraction from the simulation is a promising avenue for determining \fesclyc\ of actual observations \citep[see also][]{katz2020, katz2022, Katz2023_sphinxrelease, Choustikov2023_}. In this study, the primary limitation stems from relying on a single zoom-in simulation to forecast \fesclyc. This approach inherently restricts the range of potential \fesclyc\ values to those observed within the simulation itself, with the maximum \fesclyc\ observed in our virtual galaxy being around 42\% \citep[][]{mauerhofer2021}.

Further work is necessary to corroborate the simulation's predictions by comparing them with direct \fesclyc\ measurements from lower-redshift samples. However, we demonstrate in the following subsection that the simulation-based predictions are in alignment with the anticipated intensity of LyC leakage for each object, as inferred from their \lya\ and UV spectral properties. 

\subsection{Testing the consistency of the predicted \fesclyc}
\label{sec:lycexp}

Here we assess if the \fesclycpre\ derived in the previous section correlates with other established LyC diagnostics. 

 \subsubsection{Comparing to the \lya\ escape fractions}
 \label{sec:lyclya}

The \lya\ line is a powerful tracer for the LyC leakage \citep[e.g.,][]{verhamme2017, steidel2018, izotov2023_lyaemission_deficientmetal}, as \lya\ photons are affected by the same neutral hydrogen gas and dust extinction regulating the escape of LyC photons. \citet{Begley2023_lya_lyc} investigated the connection between the escape fraction of \lya\ (\fesclya) and \fesclyc\ within a sample of 152 \vandels\ objects, 11 of which are also part of the 93 well-matched \vandels\ galaxies. These 11 objects span M$_\star$ values from 10$^{8.8}$ to 10$^{9.8}$ M$_\odot$ and SFR values between 10$^{0.8}$ to 10$^{1.8}$ M$_\odot$ yr$^{-1}$. Further, the median $\chi^2$ of their 30 best mock spectra spans values from 0.7 to 1.9. This indicates that these 11 objects represent the broader well-matched \vandels\ sample examined in this study and do not exhibit bias toward specific groups of objects that are ``exceptionally" well-matched or possess a particular set of galaxy properties.

\begin{figure}[!htbp]
    \centering
        \includegraphics[width =\hsize]{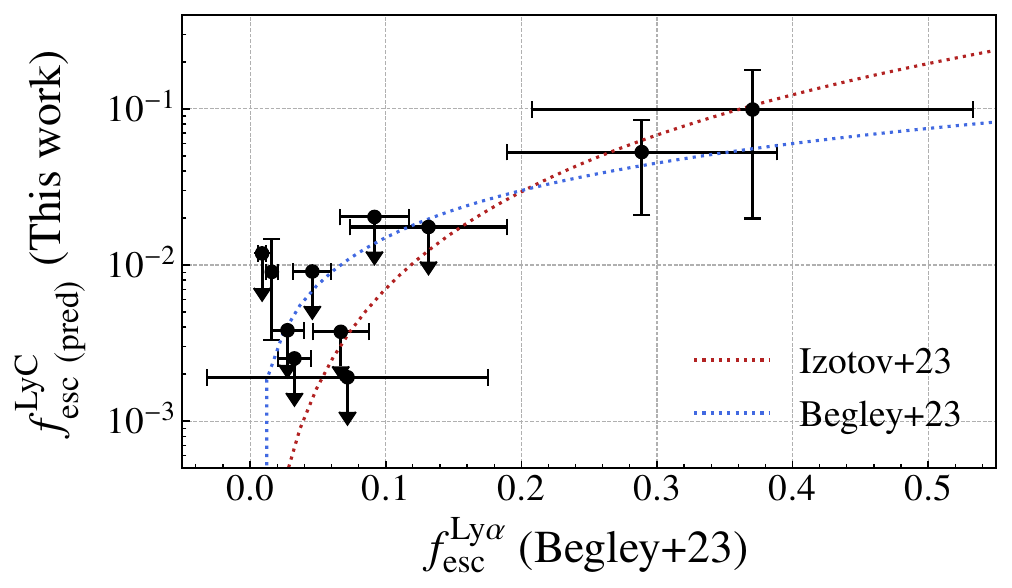}
    \caption{The comparison of the \fesclycpre\ derived using the simulation and the \lya\ escape fractions measured in \citet{Begley2023_lya_lyc} for the 11 \vandels\ objects that are present in both samples. The red and blue dotted lines show the empirical relations found in \citet{izotov2023_lyaemission_deficientmetal} (using a $z~<~0.5$ LyC emitter sample) and in \citet{Begley2023_lya_lyc} (using the \vandels\ sample), respectively.  }
    \label{fig:lyalyc}
\end{figure}

Figure~\ref{fig:lyalyc} compares the predicted LyC escape fractions and the \fesclya\ values measured in \citet{Begley2023_lya_lyc} for these objects. We observe a significant correlation between \fesclycpre\ and the measured \fesclya. Notably, this trend is highly consistent with two empirical \fesclyc--\fesclya relationships, one derived in \citet{Begley2023_lya_lyc} using \vandels\ galaxies, and the other established in \citet{izotov2023_lyaemission_deficientmetal} using a sample of $z<0.5$ LyC emitters. This consistency is remarkable as it shows that our LyC escape fraction predictions, derived from comparing LIS metal line spectra in observations and RHD simulations, adhere to established \lya-LyC observational trends. This result underscores the consistency of both \lya\ and LIS metal lines as indicators of \fesclyc. Further, it brings confidence in the \fesclycpre, suggesting these predictions may be close to the actual LyC escape fractions of these objects.

\begin{figure*}[!htbp]
    \centering
        \includegraphics[width =\hsize]{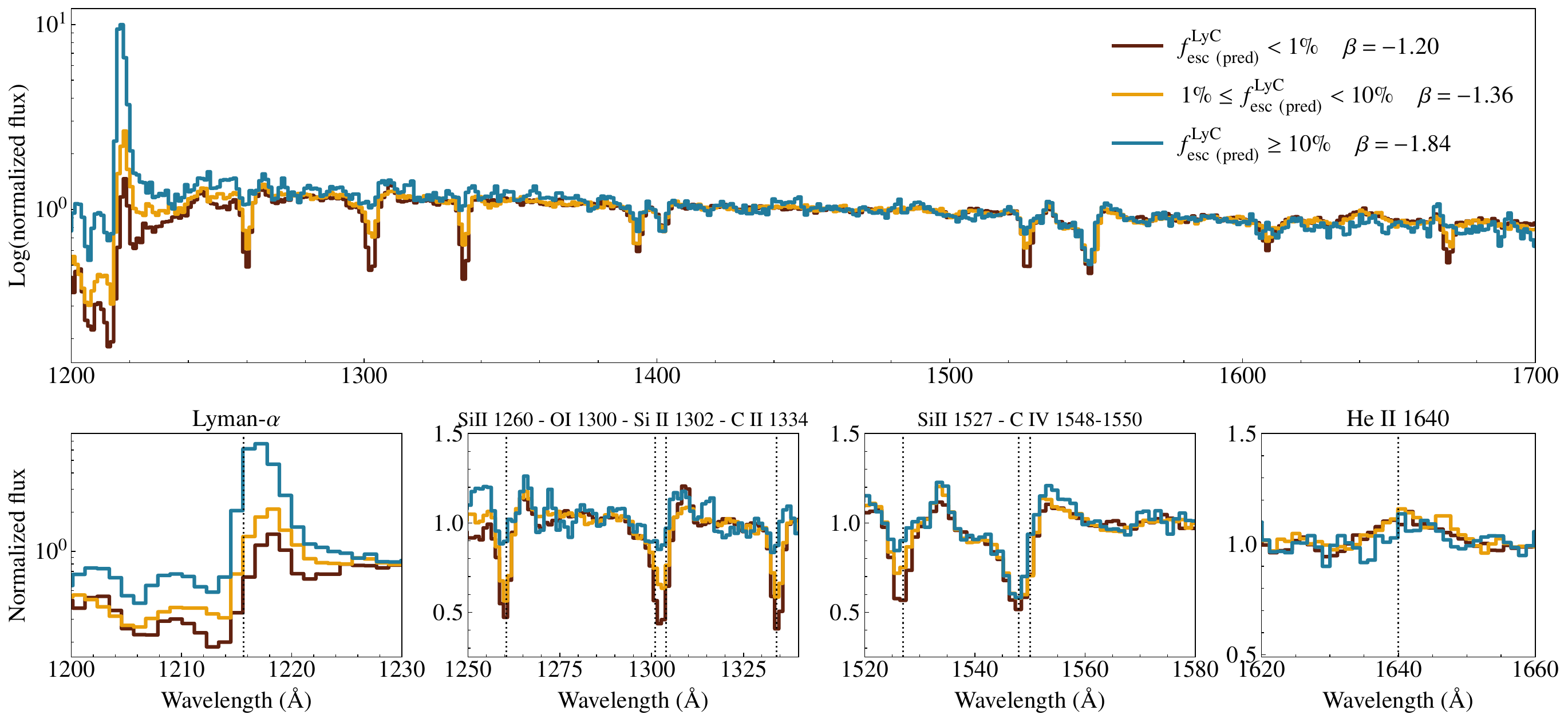}
    \caption{Composite stacks of \vandels\ spectra based on the \fesclycpre\ from this work. We use three bins of non, weak, and (relatively) strong leakers using cuts at 1 and 10\% leakage. We measure for each stack the $\beta$ slope using the observed continuum flux between 1250 and 1700\AA, and the four panels on the bottom zoom on the \lya\ profiles, LIS absorption lines, and high-energy  \ion{C}{4} and \ion{He}{2} features.}
    \label{fig:predlyc}
\end{figure*}

 \subsubsection{Comparing UV spectral differences}
\label{sec:uvdiff}

\citet{Begley2023_lya_lyc} investigated the spectral difference of composite spectra created using bins of \fesclya\ in \vandels. Here, we explore a similar exercise but producing composite stacks in three bins of \fesclycpre: below 1\%, between 1\% and 10\%, and larger or equal to 10\% respectively. We use these three \fesclycpre\ bins to represent categories of non, weak, or relatively strong leakers, respectively. As discussed in Section~\ref{sec:lycpre}, only two objects have \fesclyc\ $\geq10\%$. To increase the number of objects in those bins, we also include four galaxies whose predicted escape fractions are consistent with a strong leakage (i.e. \fesclycpre$+1\sigma>10\%$).

Figure~\ref{fig:predlyc} present these three stacks, with four panels zooming in on the \lya\ profiles, LIS absorption lines (from \ion{Si}{2}, \ion{O}{1}, and \ion{C}{2}), and the high-energy UV features \ion{C}{4} and \ion{He}{2}, tracing hard ionizing radiation field above 47.9 and 54.4 eV respectively. For each of the stacks, we measure the $\beta$ slope between 1250 and 1700\AA\ (selecting feature-free intervals in that range) by fitting a line such that $F_{\rm obs} \propto \lambda^\beta$. $\beta$ is a robust indicator of the LyC leakage in low-$z$ LyC leakers \citep{chisholm2022_beta} because the UV slope strongly depends on the dust extinction which also regulates the escape of ionizing photons. $\beta$ is especially powerful because its measurement can be relatively simply extracted from spectra of high-$z$ objects with JWST \citep[e.g.,][]{cullen2023_beta_jwst, cullen2023_dustfreepop}.

Examining the distinctions among the three composite spectra, we observe several trends. Stacks with higher predicted \fesclyc\ exhibit bluer $\beta$ values, an increase in \lya\ flux, and weaker LIS absorption lines, respectively. These observations align with the physical picture that lower levels of dust and absorbing gas facilitate and enhance the escape of both LyC and \lya\ photons \citep{gazagnes2020, Begley2023_lya_lyc}. We note that the measured $\beta$ slopes in the weak and relatively strong leakers are ``redder" than what is typically expected from low-$z$ LyC emitters. Using the $\beta$ to \fesclyc\ relation from \citet{chisholm2022_beta} derived from the LzLCS sample \citep{flury2022_lzlcs}, a $\beta$ slope of $-1.35$, and $-1.84$, would correspond to \fesclyc\ of $\sim1\%$, and $\sim3\%$, respectively, hence lower than what is expected based on the simulation-based predictions.

Regarding the high-ionization UV features, we observe no difference in the \ion{C}{4} absorption or \ion{He}{2} emission. However, we do observe a slightly larger \ion{C}{4} P-Cygni emission flux in the \fesclyc$>10\%$ stack, which could be either due to a younger stellar population \citep{chisholm2020} or enhanced collisionally-excited \ion{C}{4} emission in a high-ionization galaxy environment \citep{2berg2019_civ}. Prominent \ion{C}{4} and \ion{He}{2} emission are sometimes observed in extreme LyC leakers \citep[\fesclyc\ $>20\%$ or \fesclya$>50\%$, see][]{schaerer2022_civ, naidu2022_lya50} with large intrinsic production of ionizing photons. None of the \fesclycpre\ derived in this work are consistent with the presence of an extreme leakage \footnote{We note that \citet{saxena2020_heII} studied the properties of 51 \ion{He}{2} emitters in \vandels\ but none of these objects are present in our current sample}. We note that \citet{marques2022_spectralhardness} explored the correlation between strong LyC leakage and spectral hardness in a sample of 89 galaxies and found that extreme leakers do not have harder ionizing spectra than non-leakers at similar metallicity. Hence, this aspect may also explain the absence of differences seen for higher-energy lines in the three stacks.

Overall, this section highlights that the predicted LyC escape fractions derived from this simulation align closely with empirical LyC trends observed in observational studies, reinforcing our confidence in the physical accuracy of the current RHD simulation. This means we can use the virtual galaxy environment to elucidate the physical processes responsible for the LyC escape \citep[see also][]{katz2020,katz2022, Katz2023_sphinxrelease,Choustikov2023_}, and we explore that aspect in the next section.

\subsection{What makes LIS lines good tracers of \fesclyc}
\label{sec:lycsimu}

In the previous Section, we highlighted that the \fesclycpre\ values, extracted from the simulation from the simple comparison of the LIS line spectra in the observations and simulation, consistently scale with independent LyC diagnostics such as \lya\ and the $\beta$ slopes, and align well with direct and indirect LyC constraints established by previous studies \citep{Begley2022_lyc, saldana2023}. This result is significant as it reinforces both theoretical and empirical evidence suggesting that LIS metal line spectra are reliable indicators of LyC leakage  \citep[e.g.][]{heckman2001, alexandroff2015, gazagnes2018, saldana2023}. 

Yet, this outcome brings us to the following question: what makes the LIS metal lines good tracers of the LyC leakage? The tight \fesclyc\ empirical trends found in low-$z$ galaxies over the past suggested that LyC photons escape through relatively dust-free, low neutral gas column-density channels \citep{gazagnes2018, chisholm2018, saldana2023, reddy2016stack}. Further, \citet{flury2022_lycdiag} found a significant correlation between larger LyC leakage and the objects' compactness, which may support that stellar feedback more efficiently clear escape channels in such environment \citep{cen2020}. Hence, LIS metal lines may trace the LyC leakage because they indirectly probe the presence of compact, dust-free, and neutral-gas-free regions through which LyC photons can escape.


\begin{figure}[!htbp]
    \centering
        \includegraphics[width =\hsize]{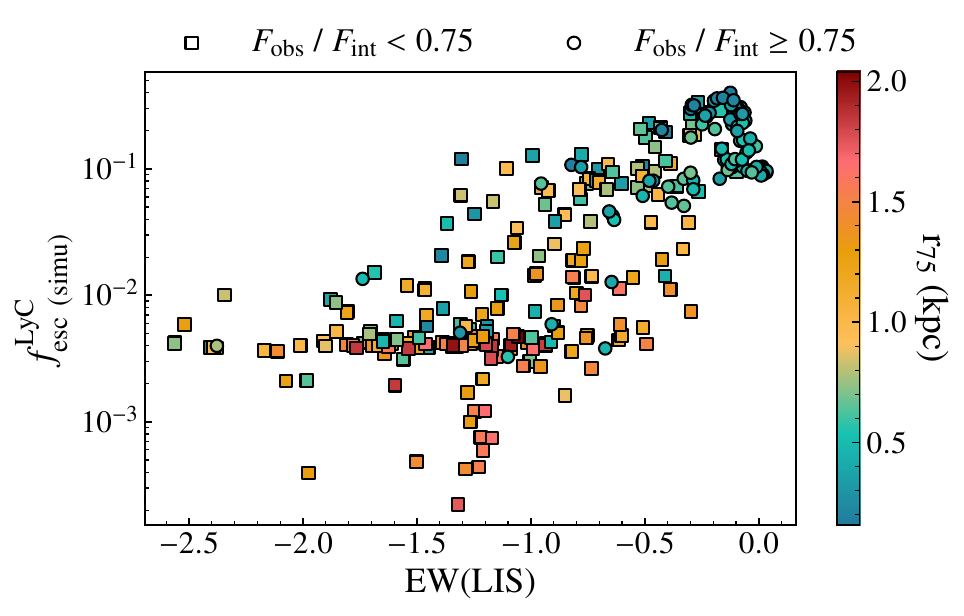}
    \caption{\fesclyc\ as a function of the average EW of the \siII\ and  \cII\ absorption lines (noted EW(LIS)) along 300 lines of sight for a single simulation output of the virtual galaxy at t $=$ 2 Gyrs. Squares and circles differentiate sightlines where less or more than 75\% of the intrinsic flux escapes the galaxy, respectively (using the luminosity weighted $F_{\rm obs}/F_{\rm int}$ ratio). Each point is color-coded by r$_{75}$, which is the radius at which the integrated flux accounts for more than 75\% of the total escaping flux in that direction. }
    \label{fig:fescsimu}
\end{figure}

We explore here the complex relation between \fesclyc, LIS metal line EWs, dust attenuation, and compactness. To do so, we generate Integral Field Unit cubes for each of the 300 lines of sight for a single simulation output of the galaxy at t $=$ 2 Gyrs (the time-step where most of the best mock spectra are found). Each data cube comprises $64\times64$ spatial pixels, with each pixel corresponding to a physical size of $120\times120$ pc, covering a total area of approximately $7.7$ by $7.7$ kpc on the entire cube face. 

For each of these 300 viewing directions, we extract three essential properties: (i) the radius at which the integrated flux accounts for more than 75\% of the total flux (referred to as r$_{75}$), (ii) the luminosity-weighted dust attenuation, calculated by comparing the observed and intrinsic flux and taking the luminosity-weighted average across all pixels, and (iii) the EW of the absorption lines for \siII\ and \cII\ (denoted as EW(LIS)) along that specific line of sight. We opted for the use of r$_{75}$ as a measure of compactness instead of the half-light radius r$_{50}$ primarily because in numerous viewing directions, a single pixel accounts for 50\% of the total flux. This makes it challenging to discern meaningful variations when relying only on r$_{50}$. 

Figure~\ref{fig:fescsimu} shows the connection between \fesclyc\ in the simulation, EW(LIS), r${75}$, and the dust attenuation. Directions with high \fesclyc\ exhibit low EW(LIS), minimal dust attenuation, and a smaller r${75}$. This aligns with the expected LyC-leakage picture: in our virtual galaxy, LyC photons likely escape through areas that are spatially confined (indicated by a low r$_{75}$), have a low neutral gas amount (low EWs of ions tracing the low-ionized gas) and are dust depleted (higher F${\rm obs}$/F$_{\rm int}$ ratio).

Interestingly, these three characteristics (lack of neutral gas and dust and spatial compactness) tend to occur together. Directions with low EW(LIS) usually have lower dust attenuation and a smaller r$_{75}$. Therefore, this explains why LIS line profiles are good tracers of the \fesclyc\ values: they indirectly reflect all the factors influencing LyC leakage. We note that this is not always true, a few directions have only one or two of these criteria met (e.g., low r$_{75}$ and dust attenuation but high EW(LIS), which results in lower \fesclyc. These cases might explain the observed scatter in LyC diagnostic trends since LyC diagnostics usually trace only one of these characteristics.

In general, Figure~\ref{fig:fescsimu} supports a physical picture where LyC photons escape through compact regions, depleted of neutral gas and dust. However, further work is needed to refine this model and to more thoroughly evaluate the reliability and the physics behind existing LyC diagnostics. This effort will be the topic of a separate paper.

\section{Summary}
\label{sec:sum}

In this study, we compared mock UV \cII\ and \siII\ absorption and emission line spectra simulated from a single $\sim10^9$ $M_\odot$ virtual galaxy to the observed spectra of 131 $z\sim3$ galaxies of stellar mass between $\sim10^9$ and $10^{10.5}$ from the \vandels\ survey \citep{McLure2018_vandels, Pentericci2018_vandels, Garilli2021_vandels}. We assessed the degree of resemblance between the simulated and observed spectra and investigated the potential of the LIS metal line spectra as tracers of galaxy properties and Lyman Continuum leakage within star-forming galaxies. In particular, we established a simulation-based \fesclyc\ estimation method which enabled us to predict the strength of the LyC leakage in the selected \vandels\ sample.  We summarize our findings as follows:

\begin{itemize}[leftmargin=*]
    \item  The correlation and dispersion in the distribution of equivalent widths for the \cII\ and \siII\ line spectra closely match the \vandels\ observations. We note a slight shift towards higher EWs in the observations, with 33\% of the \vandels\ measurements falling outside the range of simulated values.  This difference is attributable to the generally larger stellar masses of the selected \vandels\ galaxies compared to the virtual galaxy (Section~\ref{sec:linesmeasurements}).
    
    \item For each  \vandels\ galaxy, we determine the similarity ($\tilde{\chi}^2$) between the mock spectra and the observed absorption and emission line profiles. We find for 37\% of the \vandels\ spectra a most spectrum with excellent agreement ($\tilde{\chi}^2 < 1$), while 83\% are reasonably well matched (best mock spectrum with $\tilde{\chi}^2<2$). Notably, we observed a dependence on stellar mass, with the simulated spectra failing to reproduce the most massive galaxies (M$\star > 10^{10.1}$M$\odot$, Section~\ref{sec:compsfr}) due to the presence of broader absorption profiles. This finding further supports the conclusions from \citet{gazagnes2023, blaizot2023simulating}, highlighting the limitations of the current virtual galaxy in representing the environments of the most massive galaxies.

    \item The best-matching mock spectra originate from time steps when the virtual galaxy's luminosity and SFR are at or close to a peak (Section~\ref{sec:galprop}). These luminosity peaks are four times larger than during weaker star formation phases. The influence of intense star formation in explaining the good agreement between simulated and observed spectra is consistent with the nature of the selected \vandels\ galaxies as UV-bright, relatively star-forming objects. This aspect emphasizes that star-formation phases have a direct impact on the LIS metal line spectra. In general, this outcome reminds us of the strong observational biases of current surveys that disproportionately favor galaxies in a star-bursting phase at any given stellar mass (Figure~\ref{fig:origin_sfr}). 
    
    \item We predicted the \fesclyc\ for each galaxy in the selected \vandels\ sample using the virtual galaxy environment (\fesclycpre, Section~\ref{sec:lycpre}). We derived an average predicted \fesclycpre\ of 0.01 $\pm$ 0.02, within 2$\sigma$ of the value 0.07$\pm$0.02 reported by \citet{Begley2022_lyc}, and directly consistent with the indirect constraint from \citet{saldana2023}  who used the residual flux of LIS absorption lines and the expected dust attenuation at 912\AA.

     \item The simulation-based \fesclycpre\ tightly correlate with the \lya\ escape fractions from \citet{Begley2023_lya_lyc} and are in remarkable agreement with the empirical \fesclyc--\fesclya relations from \citet{izotov2023_lyaemission_deficientmetal} and \citet{Begley2023_lya_lyc} (Section~\ref{sec:lyclya}, Figure~\ref{fig:lyalyc}). We produced composite spectra in bins of \fesclycpre\ (Section~\ref{sec:uvdiff}) and show that the stack comprising the higher \fesclycpre\ galaxies exhibit bluer $\beta$ values, an increase in \lya\ flux, and weaker LIS absorption lines. We do not find significant differences in high-energy lines such as  \ion{C}{4} and \ion{He}{2}, which may suggest that the current sample does not hold extreme LyC leakage candidates or that the relation between \fesclyc\ and spectral hardness is more complex than expected.

    \item  Building upon the good agreement between the simulation-based \fesclycpre\ and well-established LyC empirical trends, we explored what makes LIS lines good tracers of  \fesclyc\ in the simulation (Section~\ref{sec:lycsimu}). We find that directions with larger \fesclyc\ exhibit lower LIS absorption line EW, lower dust attenuation, and compact spatial profiles (smaller r$_{75}$). These results align with the expected LyC-leakage picture derived from observations where LyC photons likely escape through spatially confined areas depleted of neutral gas and dust.  The consistent co-occurrence of these three characteristics (low neutral gas, low dust, and spatial compactness) may explain why LIS-based LyC diagnostics provide relatively robust indirect estimates into the \fesclyc\ of galaxies.
\end{itemize}

This study extends the success of using a zoom-in simulation approach to analyze UV absorption and emission lines, as seen in recent studies \citep{mauerhofer2021, blaizot2023simulating, gazagnes2023}. It reaffirms the potential of these methods in understanding the factors contributing to various and intricate line profiles in galaxies. Additionally, it underscores their relevance in the context of high-redshift observations \citep[see also][]{Katz2023_sphinxrelease, Choustikov2023_} and their potential to provide valuable insights, including estimates of the LyC escape fraction, in the current epoch of reionization-era observations with JWST.


\begin{acknowledgments}

The authors thank the referee for comments that improved the quality of this manuscript. SG is grateful for the support enabled by the Harlan J. Smith McDonald fellowship which made this project possible. FC acknowledges the support from a UKRI Frontier Research Guarantee Grant (PI Cullen; grant reference EP/X021025/1). TG and AV are supported by the professorship grant  SNFProf\_PP00P2\_211023 from the Swiss National Foundation for Scientific Research.  VM acknowledges support from the NWO grant 016.VIDI.189.162 (``ODIN”). This manuscript does not talk about birds.
\end{acknowledgments}

\subsection*{Data Availability}
The VANDELS survey is a European Southern Observatory Public Spectroscopic Survey. The full spectroscopic data set, together with photometric catalogs and derived quantities, is available from http://vandels.inaf.it, as well as from the ESO archive https://www.eso.org/qi. Any secondary product and/or data underlying this article will be shared on reasonable request to the corresponding author."

\software{Astropy \citep{astropy:2013, astropy:2018, astropy:2022}, NumPy \citep{numpy}, pandas \citep{panda}, SciPy \citep{scipy}, RASCAS \citep{rascas2009}}

%






\appendix

\section{Accounting for a redshift uncertainty when finding the best-matching spectra}
\label{app:offseteffect}

In Section~\ref{sec:fitting}, we implemented a matching strategy to find, for each \vandels\ spectrum analyzed in this work, the best-matching mock spectra in the simulation. We discussed that, although the selected \vandels\ objects have a redshift reliable at $>95\%$, a few of these redshift estimates rely on ISM features that can be kinematics-dependent and therefore result in slight offsets (typically within a few hundred km s$^{-1}$ at most) compared to the systemic redshift typically estimated using optical nebular emission lines. In our main analysis, we addressed this redshift uncertainty by initially conducting a cross-correlation between the observed spectra and each mock spectrum to determine the optimal spectral shift required to align the spectral features. Yet we noted that this approach nullifies actual differences in the line velocity offsets between the mock and observed spectra.

\begin{figure}[!htbp]
    \centering
        \includegraphics[width=\hsize]{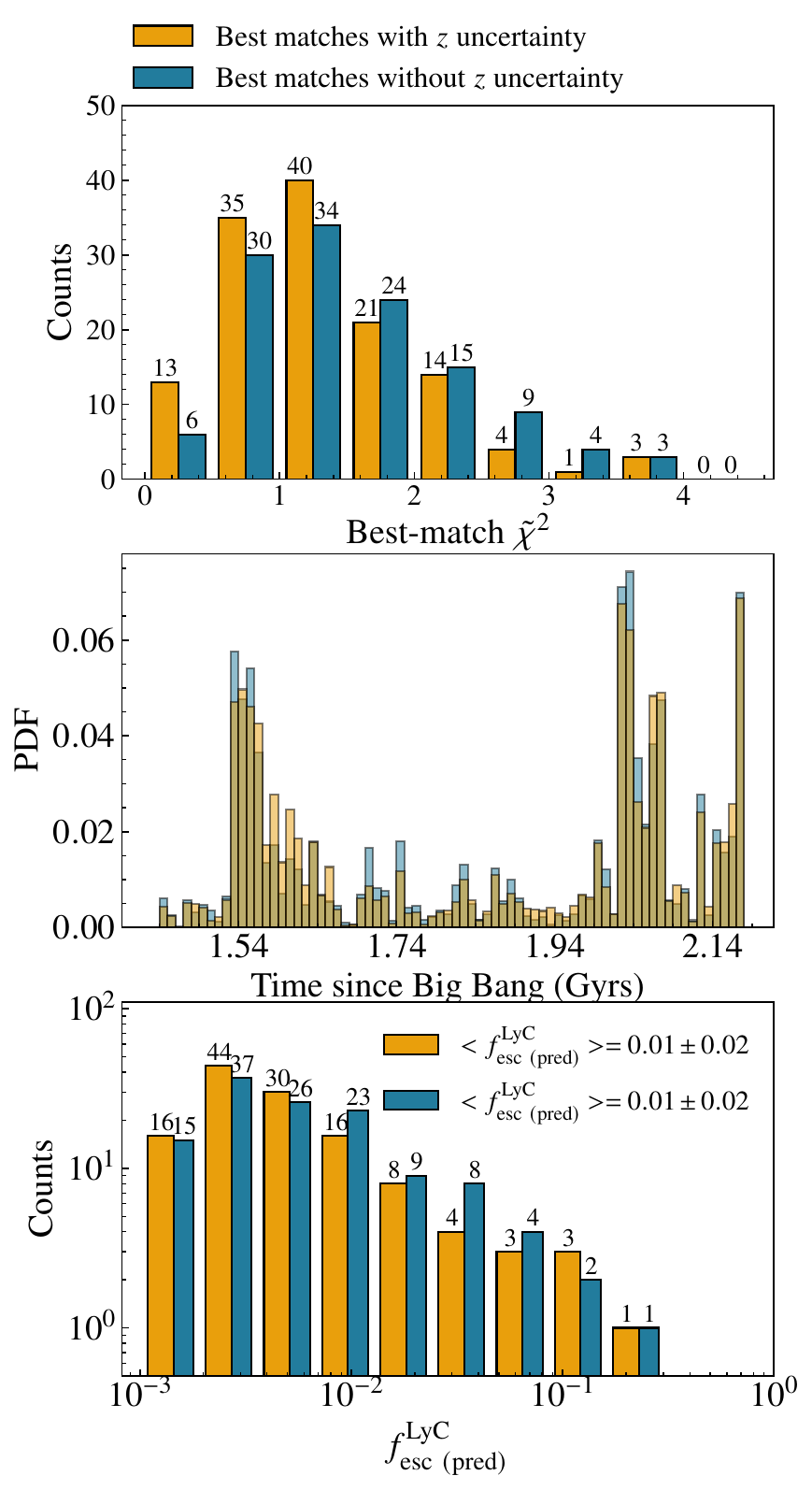}
    \caption{Exploring the impact of accounting for potential redshift uncertainties when finding the best-matching mock spectra. Top: The comparison of the best-match $\tilde{\chi}^2$ for the two different cases. As expected, allowing for a spectral shift when finding the best matching spectra leads to typically better $\tilde{\chi}^2$. Middle: The differences in the distribution of the best-matching spectra with and without $z$ uncertainty. We do not find a significant difference in both distributions. Right: Comparing the predicted \fesclyc\ when using the two different approaches. We do observe small variations of the \fesclycpre, however, these variations are within the uncertainties and do not change the interpretation of galaxies which could be categorized as non, weak, or strong leakers. Overall, these there panels support that choosing one or the other approach has no impact on the main conclusions drawn in this work. }
    \label{fig:app}
\end{figure}

To assess the potential impact of this pre-alignment step, we conducted an independent experiment, detailed here, where we identified the best-matching mock spectra both with and without including this ``redshift uncertainty". In Figure~\ref{fig:app}, we present the main differences of using either approach. The top panel shows the distribution of $\tilde{\chi}^2$ values and highlights that a few more galaxies have $\tilde{\chi}^2 <1.5$ when we first pre-align the mock and observed spectra. This result is not unexpected as by performing a pre-alignment, we already find the spectral shift that provides the minimal difference between both spectra.
The middle panel reproduces Figure~\ref{fig:orig_extract} from Section~\ref{sec:spec_ori} and compares whether we observe any differences in the time-steps origin of the 30 best-matching spectra for each \vandels\ object when we do or do not account for potential redshift uncertainties. The two distributions are highly similar, suggesting that the conclusions drawn in Section~\ref{sec:spec_ori} are unaffected by the choice of either method. 
Finally, the bottom panel compares the differences in the predicted \fesclyc, using the methodology detailed in Section~\ref{sec:lyc}, when one or the other approach. The predicted values are relatively consistent with each other and choosing one or the other approach does not impact the overall predicted average LyC escape fraction (0.01$\pm$0.02 in both cases). 

 Overall, Figure~\ref{fig:app} emphasizes that choosing a best-matching approach that does not include a pre-alignment step would have had no impact on the main conclusions drawn in this work.

\end{document}